\documentclass[aps,prl,showpacs,twocolumn,superscriptaddress]{revtex4}
\usepackage{bm,color}
\usepackage{graphicx}
\usepackage{amsmath, amssymb}
\usepackage{color}
\usepackage{braket}
\usepackage{comment}

\begin{document}
\title {Floquet theory of photoinduced topological phase transitions in the organic salt $\alpha$-(BEDT-TTF)$_2$I$_3$ irradiated with elliptically polarized light}
\author{Keisuke Kitayama}
\affiliation{Department of Physics, University of Tokyo, Hongo, Bunkyo-ku, Tokyo 113-8656, Japan}
\author{Yasuhiro Tanaka}
\affiliation{Department of Applied Physics, Waseda University, Okubo, Shinjuku-ku, Tokyo 169-8555, Japan}	
\author{Masao Ogata}
\affiliation{Department of Physics, University of Tokyo, Hongo, Bunkyo-ku, Tokyo 113-8656, Japan}
\affiliation{Trans-scale Quantum Science Institute, University of Tokyo, Bunkyo-ku, Tokyo 113-0033, Japan}
\author{Masahito Mochizuki}
\affiliation{Department of Applied Physics, Waseda University, Okubo, Shinjuku-ku, Tokyo 169-8555, Japan}
\begin{abstract}
We theoretically investigate possible photoinduced topological phase transitions in the organic salt $\alpha$-(BEDT-TTF)$_2$I$_3$, which possesses a pair of inclined massless Dirac-cone bands between the conduction and valence bands under uniaxial pressure. The Floquet analyses of a driven tight-binding model for this material reveal rich photoinduced variations of band structures, Chern numbers, and Hall conductivities under irradiation with elliptically polarized light. The obtained phase diagrams contain a variety of nonequilibrium steady phases, e.g., the Floquet Chern insulator, Floquet semimetal, and Floquet normal insulator phases. 
This work widens a scope of target materials for research on photoinduced topological phase transitions and contributes to development of research on the optical manipulations of electronic states in matters.
\end{abstract}
\maketitle

\section{1. Introduction}
\begin{figure}[htb]
\includegraphics[scale=0.5]{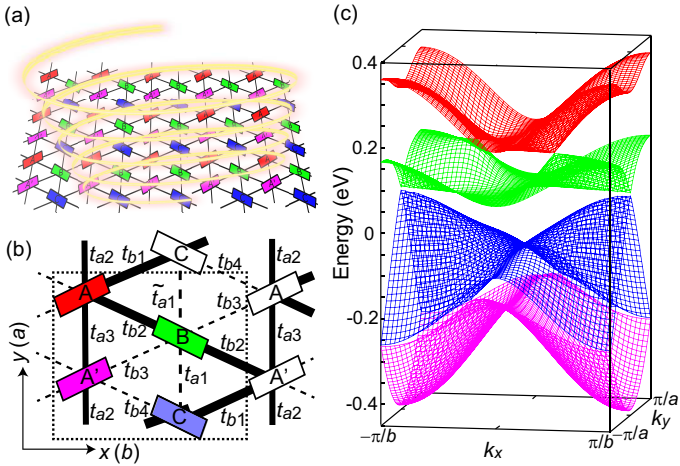}
\caption{(a) Schematic of $\alpha$-(BEDT-TTF)$_2$I$_3$ irradiated with elliptically polarized light. (b) Two-dimensional (BEDT-TTF)-layer in $\alpha$-(BEDT-TTF)$_2$I$_3$ with a unit cell composed of four molecular sublattices of A, A$^{\prime}$, B and C and eight kinds of transfer integrals for a tight-binding model of the BEDT-TTF layer. Dashed rectangle represents the unit cell with $b$ and $a$ being the lattice constant along $x$ and $y$ direction, respectively. (c) Band dispersion relations of $\alpha$-(BEDT-TTF)$_2$I$_3$ obtained by diagonalizing the Hamiltonian matrix of the tight-binding model in Eq.~(\ref{eq:alpham}). There exist a pair of inclined Dirac-cone bands between the third and fourth bands, where the Dirac points are located on the Fermi level.}
\label{Fig01}
\end{figure}
Photoinduced phase transitions have attracted a great deal of research interest~\cite{Yonemitsu06,Tokura06,Aoki14,Bukov15,Basov17}. The Floquet theory has been widely exploited for theoretical studies on nonequilibrium steady states and their phase transitions in photodriven systems~\cite{Oka09,Kitagawa10,Kitagawa11,Lindner11,Ezawa13,Grushin14,ZhengW14,Takasan15,Mikami16,JYZou16,Claassen16,ZYan16,LDu17,Ezawa17,Takasan17a,Takasan17b,Menon18,ChenR18,ZhangMY19,Kang20,Tanaka20,Kitayama20,Sato16,Takayoshi14,Kitamura17,Ikeda20}. Emergence of quantum Hall effects in a tight-binding model on the honeycomb lattice irradiated with circularly polarized light was theoretically predicted using the Floquet theory~\cite{Oka09,Kitagawa11}, in which the band structure attains topological nature similar to that described by the Haldane model~\cite{Haldane88}. This topological phase transition manifests photoinduced quantum Hall effects without external magnetic fields, which was indeed observed by recent optical experiments in graphene~\cite{McIver19}. 

Since these pioneering works, many studies based on the Floquet theory have been performed for both electron systems~\cite{Oka09,Kitagawa10,Kitagawa11,Lindner11,Ezawa13,Grushin14,ZhengW14,Takasan15,Mikami16,JYZou16,Claassen16,ZYan16,LDu17,Ezawa17,Takasan17a,Takasan17b,Menon18,ChenR18,ZhangMY19,Kang20,Tanaka20,Kitayama20} and spin systems~\cite{Sato16,Takayoshi14,Kitamura17,Ikeda20}. However, most of these studies dealt with toy models on simple lattices or simple two-dimensional materials such as graphene~\cite{Oka09,Kitagawa11,Kitagawa10}, silicene~\cite{Ezawa13}, stacked graphene~\cite{JYZou16}, transition-metal dichalcogenides~\cite{Claassen16,ZhangMY19}, and black phosphorene~\cite{Kang20}. On the other hand, bulk materials have scarcely been involved in a scope of this kind of research. However, for further development of this promising research field, it is indispensable to widen the range of target materials, and, for this sake, theoretical studies on real bulk materials with complex crystal and electronic structures are highly desired. In the research on real bulk materials, we expect richer material-specific photoinduced topological phenomena and physics.

The organic compound $\alpha$-(BEDT-TTF)$_2$I$_3$ is one of the promising candidate materials for research in this direction~\cite{Tajima06}. Conduction electrons confined in a BEDT-TTF layer form a pair of inclined Dirac cones in the band structure when a uniaxial pressure $P_a>2$ kbar is applied along the $a$ axis~\cite{Katayama06,Kobayashi07,Kajita14}. Under an ambient pressure, this material exhibits a charge-ordered ground state and has a band gap at the Fermi level. This charge order melts above the critical temperature of $T_{\rm CO}$$\sim$135 K, and a zero-gap semiconducting state with Dirac-cone bands appears even without uniaxial pressure. A lot of interesting topologically nontrivial properties and phenomena due to the Dirac-cone bands have been theoretically proposed for this organic compound to date, e.g., the quantum Hall effect~\cite{Kajita14}, the structures of Berry curvature in momentum space~\cite{Suzumura11}, and the flux-induced Chern insulator phases~\cite{Osada17}. 

In this paper, we theoretically study possible photoinduced topological phases in organic compound $\alpha$-(BEDT-TTF)$_2$I$_3$ with a pair of inclined Dirac-cone bands. The Floquet analyses of a driven tight-binding model for this material reveals that the irradiation with elliptically polarized light opens a gap at the Dirac points located on the Fermi level, and the system becomes a topological insulator. This so-called Floquet Chern insulator phase as a photoinduced nonequilibrium steady phase is characterized by a quantized topological number~\cite{Thouless82,Kohmoto85} and conducting chiral edge states~\cite{Hao08}. Moreover, we obtain rich phase diagrams as functions of amplitudes, frequency and ellipticity of the light, which involve Chern insulator, semimetal, and normal insulator phases as nonequilibrium steady states. We also discuss the behaviors of Hall conductivity associated with the photoinduced phase evolution. In fact, the photoinduced phase transitions for this compound have been theoretically studied in the case of circularly polarized light~\cite{Kitayama20}. In the case of elliptically polarized light studied in the present paper, we obtain richer phase diagrams and phase evolutions as functions of the light ellipticity. Moreover, we discuss edge states and the Hall conductivities in the photoinduced states in detail.

As argued in our previous paper~\cite{Kitayama20}, one advantage of the usages of organic compounds is that an effective amplitude of light is an order of magnitude larger than graphene because of much larger lattice constants as discussed later, which enhances experimental feasibility of the predicted photoinduced topological phase transitions. Another advantage of $\alpha$-(BEDT-TTF)$_2$I$_3$ is that the bands near the Fermi level [i.e., four bands in Fig.~\ref{Fig01}(c)] relevant to the electronic properties of this material are well separated from bands above and below them. This specific band structure provides us a precious opportunity to have a finite window of light frequency that satisfies the so-called off-resonant condition, with which the Floquet bands of different photon numbers do not overlap with each other and, thereby, the nonequilibrium steady phases are well-defined. We will discuss these aspects later in more detail.

In the following, we formulate the Floquet theory to analyze a time-periodic Hamiltonian in Sect.~2 and then apply this formalism to the photodriven tight-binding model for $\alpha$-(BEDT-TTF)$_2$I$_3$ irradiated with elliptically polarized light in Sect.~3. In Sect.~4, we explain how to calculate the physical quantities such as Chern numbers, Hall conductivities, and band structures associated with edge states within the Floquet formalism. In Sect.~5, we show and discuss the results for photoinduced variations of band structurers, Chern numbers, and Hall conductivities as well as the phase diagrams. We also discuss the chiral edge states in the Floquet Chern insulator phase, which manifest the topological nature of this photoinduced insulating phase. Finally, we summarize this work in Sect.~6.

\section{2. Floquet Theory for Photodriven Systems}
In this section, we introduce a general formalism of the Floquet theory for photo-irradiated systems. The photodriven systems are described by the time dependent Schr\"{o}dinger equation,
\begin{eqnarray}
i\hbar\frac{\partial}{\partial \tau}\ket{\Psi(\tau)}=H(\tau)\ket{\Psi(\tau)}.
\label{eq:FLQ1}
\end{eqnarray}
Here the Hamiltonian $H(\tau)$ is time-periodic as $H(\tau)=H(\tau+T)$ with a temporal period $T(=2\pi/\omega)$ of light. In such systems, the wavefunction $\ket{\Psi(\tau)}$ can be written in the form,
\begin{eqnarray}
\ket{\Psi(\tau)} = e^{-i\varepsilon \tau/\hbar}\ket{\Phi(\tau)},
\label{eq:FLQ2}
\end{eqnarray}
where $\ket{\Phi(\tau)}$ and $\varepsilon$ are referred to as the Floquet state and the quasienergy, respectively. The Floquet state satisfies $\ket{\Phi(\tau)}=\ket{\Phi(\tau+T)}$. This theorem is called Floquet theorem and can be regarded as a temporal version of the Bloch theorem for spatially periodic systems. We substitute Eq.~(\ref{eq:FLQ2}) into Eq.~(\ref{eq:FLQ1}) and perform the Fourier transformations. Eventually, the time dependent Schr\"{o}dinger equation is rewritten in the form,
\begin{eqnarray}
\sum_{m=-\infty}^\infty \mathcal{H}_{nm} \ket{\Phi_{\nu}^m}
= \varepsilon^n_{\nu}\ket{\Phi_{\nu}^n},
\label{eq:H-Mw}
\end{eqnarray}
where 
\begin{eqnarray}
\mathcal{H}_{nm}=H_{n-m}-m\omega\delta_{n,m}.
\label{eq:H-Mw2}
\end{eqnarray}
Here the Fourier coefficients $H_n$ and $\ket{\Phi_{\nu}^n}$ are defined as
\begin{eqnarray}
H_n&=&\frac{1}{T}\int_0^T H(\tau)e^{in\omega\tau} d\tau,
\label{eq:HamFC}\\
\ket{\Phi_{\nu}^n}&=&\frac{1}{T}\int_0^T \ket{\Phi_{\nu}(\tau)}e^{in\omega\tau} d\tau.
\label{eq:phiFC}
\end{eqnarray}
The integer $n$ corresponds to the number of photons, and $\ket{\Phi_{\nu}^{-n}}$ is regarded as a state dressed with $n$ photons. After the projection to the zero-photon subspace, we obtain the effective Hamiltonian $H_\mathrm{eff}$~\cite{Mikami16},
\begin{eqnarray}
H_\mathrm{eff}=H_0 
+ \sum_{n=1}^\infty
\frac{[H_{-n}, H_n]}{n\hbar\omega} + O\left(\frac{t^3}{\hbar^2\omega^2}\right).
\label{eq:Heff}
\end{eqnarray}
Here $t$ denotes a typical value of the transfer integrals. This effective Hamiltonian is valid when the light frequency $\omega$ is high enough. It is noted that the following approximated form, in which only the first term of the series expansion is taken, is often used in the high-frequency limit,
\begin{eqnarray}
H_\mathrm{eff} \sim H_0 
+ \frac{[H_{-1}, H_1]}{\hbar\omega}.
\label{eq:Heff2}
\end{eqnarray}

Now we consider the following tight-binding model that describes a lattice electron system,
\begin{eqnarray}
H=\sum_{i,j} t_{ij} c^\dagger_i c_j,
\end{eqnarray}
where $i$ and $j$ denote the lattice sites. The symbol $c^\dagger_i$ ($c_i$) denotes the creation (annihilation) operator of an electron at $i$th site, while $t_{ij}$ represents the transfer integral between $i$th and $j$th sites. When the system is irradiated with light, the transfer integrals attain Peierls phases due to the time-dependent vector potential of light electromagnetic field. A general form of the vector potential caused by the elliptically polarized light is given by,
\begin{eqnarray}
\bm{A}(\tau)=(A_x\sin(\omega\tau+\phi), A_y\sin(\omega\tau)).
\label{eq:VecP1}
\end{eqnarray}
This vector potential gives a time-dependent electric field
\begin{eqnarray}
\bm E(\tau)&=&-\frac{d \bm{A}(\tau)}{d \tau}
\nonumber  \\
&=&-(A_x\omega\cos(\omega\tau+\phi), A_y\omega\cos(\omega\tau)).
\end{eqnarray}
The transfer integrals with a time-dependent Peierls phase are given by
\begin{eqnarray}
t_{ij}(\tau)
&=&t_{ij}\exp\left[
-i\frac{e}{\hbar} \bm{A}(\tau)\cdot(\bm{r}_{i}-\bm{r}_{j})\right]
\nonumber \\
&=&t_{ij}\exp \left[
 -i\frac{e}{\hbar}A_x(x_i-x_j)\sin(\omega\tau+\phi) \right.
\nonumber \\
& &\hspace{1.5cm} \left.
-i\frac{e}{\hbar}A_y(y_i-y_j)\sin(\omega\tau) \right].
\end{eqnarray}
Here we introduce the coordinates $\bm r_i=(x_i, y_i)$ for $i$th site.
The Fourier coefficients $H_n$ are calculated using Eq.~(\ref{eq:HamFC}) as
\begin{eqnarray}
H_n=\sum_{i,j}t_{ij}J_n(\mathcal{A}_{ij})e^{-in\theta_{ij}}c_i^\dagger c_j,
\label{eq:eff3}
\end{eqnarray}
where $J_n$ is the $n$th Bessel function, and $\mathcal{A}_{ij}$ and $\theta_{ij}$ are respectively defined as
\begin{align}
&\mathcal{A}_{ij}=\frac{e}{\hbar}[A_x^2(x_i-x_j)^2+A_y^2(y_i-y_j)^2
\nonumber \\
& \hspace{2cm }+2A_xA_y(x_i-x_j)(y_i-y_j)\cos\phi]^{1/2}
\\
&\theta_{ij}=\tan^{-1}\left[\frac{A_x(x_i-x_j)\sin\phi}
{A_x(x_i-x_j)\cos\phi+A_y(y_i-y_j)}\right].
\end{align}
The second term of $H_{\rm eff}$ in Eq.~(\ref{eq:Heff}) is calculated as
\begin{align}
&\sum_{n=1}^\infty \frac{[H_{-n},H_{n}]}{n\hbar\omega}
\nonumber \\
&=\sum_{\left<\left<i,k\right>\right>}
\sum_{n=1}^\infty \frac{2i}{n\hbar\omega}\sum_j 
\mathrm{Im}(\chi_{ij}^{-n}\chi_{jk}^n)(c^\dagger_i c_k- c^\dagger_k c_i).
\label{eq:eff4}
\end{align}
Here we define
\begin{align}
\chi_{ij}^n \equiv t_{ij}J_n\left(\mathcal{A}_{ij}\right)e^{-in\theta_{ij}}.
\end{align}
This term describes indirect electron hoppings between $i$th and $k$th sites via the in-between $j$th site connected by two transfer integrals $t_{ij}$ and $t_{jk}$. On the other hand, the first term $H_0$ of $H_{\rm eff}$ in Eq.~(\ref{eq:Heff}) is given by,
\begin{eqnarray}
H_0=\sum_{i,j}t_{ij}J_0(\mathcal{A}_{ij})c_i^\dagger c_j.
\label{eq:eff4}
\end{eqnarray}
This term describes the direct electron hoppings between $i$th and $j$th sites connected by a single transfer integral $t_{ij}$, which are renormalized by the Bessel function $J_0(\mathcal{A}_{ij})$.

\section{3. Floquet Theory for the Photodriven $\alpha$-(BEDT-TTF)$_2$I$_3$}
We employ a tight-binding model to describe the electronic structure of the BEDT-TTF layer in $\alpha$-(BEDT-TTF)$_2$I$_3$:
\begin{eqnarray}
H=\sum_{i,j}^{N} \sum_{\alpha, \beta} t_{i\alpha,j\beta} c^\dagger_{i\alpha}c_{j\beta}.
\label{eq:alpha-tbm}
\end{eqnarray}
Here $i$ and $j$ denote the unit cells, while $\alpha$ and $\beta$ denote the molecule sites A, A$^{\prime}$, B, and C. The symbol $c^\dagger_{i\alpha}$ ($c_{i\alpha}$) denotes the creation (annihilation) operator of an electron at molecule site ($i$, $\alpha$), while $t_{i\alpha,j\beta}$ represents the transfer integral between the ($i,\alpha$) and ($j,\beta$) sites. 

This compound exhibits a charge-ordered ground state at ambient pressure. When a uniaxial pressure larger than 2 kbar is applied along the $a$ axis, this charge order melts, and eventually the peculiar band structure with a pair of inclined Dirac cones emerges. In this study, we consider the situation that the uniaxial pressure of $P_a$=4 kbar is applied. The applied pressure modulates the transfer integrals $t_{i\alpha,j\beta}$ in Eq.~(\ref{eq:alpha-tbm}) which are given by eight kinds of transfer integrals shown in Fig.~\ref{Fig01} (b). The $P_a$ dependencies of the eight transfer integrals are deduced theoretically by interpolation as~\cite{Kobayashi04},
\begin{align}
&t_{a1} = \tilde{t}_{a1} = -0.028(1.00+0.089P_a)\;{\rm eV}, \label{eq:ta1}\\
&t_{a2} = 0.048(1.0 + 0.167P_a)\;{\rm eV},\\
&t_{a3} = -0.020(1.0 - 0.025P_a)\;{\rm eV},\\
&t_{b1} = 0.123\;{\rm eV},\\
&t_{b2} = 0.140(1.0+ 0.011P_a)\;{\rm eV},\\
&t_{b3} =-0.062(1.0+0.032P_a)\;{\rm eV},\\
&t_{b4} =-0.025\;{\rm eV}. \label{eq:tb4}
\end{align}
For $P_a$=4 kbar, the values are
$t_{a1} = \tilde{t}_{a1} =-0.038$ eV, 
$t_{a2}= 0.080$ eV, 
$t_{a3}=-0.018$ eV, 
$t_{b1}= 0.123$ eV, 
$t_{b2}= 0.146$ eV, 
$t_{b3}=-0.070$ eV, and 
$t_{b4}=-0.025$ eV.

Using the Fourier transforms,
\begin{eqnarray}
c_{i\alpha}^{\dagger}=\frac{1}{\sqrt{N}}\sum_{\bm k}c^\dagger_{\bm k\alpha} 
e^{i\bm k\cdot \bm{r}_{i\alpha}},
\end{eqnarray}
we rewrite the tight-binding Hamiltonian in the momentum representation,
\begin{eqnarray}
H=\sum_{\bm{k}}(c^{\dagger}_{\bm{k}A}\, c^{\dagger}_{\bm{k}A^{\prime}}\,c^{\dagger}_{\bm{k}B}\,c^{\dagger}_{\bm{k}C})\hat{H}(\bm{k})\left(
\begin{array}{c}
c_{\bm{k}A} \\
c_{\bm{k}A^{\prime}} \\
c_{\bm{k}B} \\
c_{\bm{k}C}
\end{array}
\right),
\end{eqnarray}
with
\begin{eqnarray}
\hat{H}(\bm{k})=\left(
\begin{array}{cccc}
0 & A_2(\bm{k}) & B_2(\bm{k}) & B_1(\bm{k}) \\
A_2^{*}(\bm{k}) & 0 & B_2^{*}(\bm{k}) & B_1^{*}(\bm{k}) \\
B_2^{*}(\bm{k}) & B_2(\bm{k}) & 0 & A_1(\bm{k}) \\
B_1^{*}(\bm{k}) & B_1(\bm{k}) & A_1(\bm{k}) & 0
\end{array}
\right),
\label{eq:alpham}
\end{eqnarray}
where
\begin{align}
&A_1(\bm{k})=2t_{a1}\cos\left(\frac{k_ya}{2}\right),\\
&A_2(\bm{k})=t_{a2}e^{ik_ya/2}+t_{a3}e^{-ik_ya/2},\\
&B_1(\bm{k})=t_{b1}e^{i(k_xb/2+k_ya/4)}+t_{b4}e^{-i(k_xb/2-k_ya/4)},\\
&B_2(\bm{k})=t_{b2}e^{i(k_xb/2-k_ya/4)}+t_{b3}e^{-i(k_xb/2+k_ya/4)}.
\end{align}
Here $b$ and $a$ are the lattice constants along the $x$ and $y$ axes, respectively [see Fig.~\ref{Fig01}(b)]. In the following calculations, we use experimental values of the lattice constants $a$=0.9187 nm and $b$=1.0793 nm~\cite{Mori12}. By diagonalizing the Hamiltonian matrix in Eq.~(\ref{eq:alpham}), we obtain the band dispersion relations shown in Fig.~\ref{Fig01}(c), which have a pair of inclined massless Dirac cones between the third and fourth bands. Because the electron filling of this compound is $n_e=3/4$, the Fermi level is located between these two bands, and the Dirac points are on the Fermi level.

We consider a situation that this system is irradiated with elliptically polarized light described by Eq.~(\ref{eq:VecP1}). 
Here we introduce dimensionless quantities $\mathcal{A}_b$ and $\mathcal{A}_a$ as,
\begin{eqnarray}
\mathcal{A}_a=\frac{eaA_y}{\hbar},
\quad\quad
\mathcal{A}_b=\frac{ebA_x}{\hbar}.
\label{eq:dlAaAb}
\end{eqnarray}
Amplitudes of the two electric-field components $E_a^\omega$ and $E_b^\omega$ for the elliptically polarized light are given by,
\begin{eqnarray}
E_a^\omega=A_y\omega=\frac{\mathcal{A}_a\hbar\omega}{ea}, 
\quad\quad
E_b^\omega=A_x\omega=\frac{\mathcal{A}_b\hbar\omega}{eb}.
\end{eqnarray}
For the Hamiltonian in the momentum representation, the effects of irradiation with elliptically polarized light can be treated simply by replacing the momenta $k_x$ and $k_y$ as
\begin{align}
&k_x \quad \rightarrow \quad k_x+\mathcal{A}_b\sin(\omega\tau+\phi),
\\
&k_y \quad \rightarrow \quad k_y+\mathcal{A}_a\sin(\omega\tau).
\end{align}

It is straightforward to obtain the Fourier coefficients of the Hamiltonian $H_n$ for $\alpha$-(BEDT-TTF)$_2$I$_3$ as in Eq.~(\ref{eq:eff3}). The transfer integrals $t_{i\alpha,j\beta}$ in Eqs.~(\ref{eq:ta1})-(\ref{eq:tb4}) are renormalized as
\begin{align}
	&t_{a1} \quad \rightarrow \quad t_{a1}J_{n}(\mathcal{A}_a/2),
	\\
	&\tilde{t}_{a1} \quad \rightarrow \quad \tilde{t}_{a1}J_{-n}(\mathcal{A}_a/2),
	\\
	&t_{a2} \quad \rightarrow \quad t_{a2}J_{-n}(\mathcal{A}_a/2),
	\\
	&t_{a3} \quad \rightarrow \quad t_{a3}J_{n}(\mathcal{A}_a/2),
	\\
	&t_{b1} \quad \rightarrow \quad t_{b1}J_{-n}(\mathcal{A}_+)e^{-in\theta_+},
	\\
	&t_{b2} \quad \rightarrow \quad t_{b2}J_{n}(\mathcal{A}_-)e^{+in\theta_-},
	\\
	&t_{b3} \quad \rightarrow \quad t_{b3}J_{n}(\mathcal{A}_+)e^{-in\theta_+},
	\\
	&t_{b4} \quad \rightarrow \quad t_{b4}J_{-n}(\mathcal{A}_-)e^{+in\theta_-}.
\end{align}
where
\begin{eqnarray}
\mathcal{A}_{\pm}=\frac{1}{4}\sqrt{4\mathcal{A}_b^2+\mathcal{A}_a^2
	\pm 4\mathcal{A}_a\mathcal{A}_b\cos\phi},
\\
\theta_\pm=\tan^{-1}\left(\frac{2\mathcal{A}_b\sin\phi}
{\pm 2\mathcal{A}_b\cos\phi+\mathcal{A}_a}\right).
\end{eqnarray}
By using the renormalizations, the Fourier coefficients of the Hamiltonian $H_n$ is explicitly given in the matrix form,
\begin{eqnarray}
\hat{H}_n(\bm k)=\left(
\begin{array}{cccc}
0 & A_{2,n}(\bm k) & B_{2,n}(\bm k) & B_{1,n}(\bm k) \\
A_{2,-n}^{*}(\bm k) & 0 & B_{2,-n}^{*}(\bm k) & B_{1,-n}^{*}(\bm k) \\
B_{2,-n}^{*}(\bm k) & B_{2,n}(\bm k) & 0 & A_{1,n}(\bm k) \\
B_{1,-n}^{*}(\bm k) & B_{1,n}(\bm k) & A_{1,-n}^{*}(\bm k) & 0
\end{array}
\right)
\nonumber \\
\end{eqnarray}
with
\begin{align}
&A_{1,n}(\bm k)=
\tilde{t}_{a1}\,e^{ik_ya/2}J_{-n}(\mathcal{A}_a/2)+t_{a1}\,e^{-ik_ya/2}J_{n}(\mathcal{A}_a/2),
\nonumber \\ \\
&A_{2,n}(\bm k)=
t_{a2}\,e^{ik_ya/2}J_{-n}(\mathcal{A}_a/2)+t_{a3}\,e^{-ik_ya/2}J_{n}(\mathcal{A}_a/2),
\nonumber \\ \\
&B_{1,n}(\bm k)=
 t_{b1}\,e^{i(k_xb/2+k_ya/4)}J_{-n}(\mathcal{A}_+)e^{-in\theta_+}
\nonumber \\ & \quad\quad
+t_{b4}\,e^{-i(k_xb/2-k_ya/4)}J_{-n}(\mathcal{A}_-)e^{+in\theta_-},
\\
&B_{2,n}(\bm k)=
 t_{b2}\,e^{i(k_xb/2-k_ya/4)}J_{n}(\mathcal{A}_-)e^{+in\theta_-}
\nonumber \\ & \quad\quad
+t_{b3}\,e^{-i(k_xb/2+k_ya/4)}J_{n}(\mathcal{A}_+)e^{-in\theta_+}.
\end{align}

In the present study, we examine the case of $\phi=\pi/2$, for which we have 
\begin{eqnarray}
\mathcal{A}_+=\mathcal{A}_-=\frac{1}{4}\sqrt{4\mathcal{A}_b^2+\mathcal{A}_a^2} \equiv \mathcal{A},
\\
\theta_+=\theta_-=\tan^{-1}\left(\frac{2\mathcal{A}_b}{\mathcal{A}_a}\right) \equiv \theta.
\end{eqnarray}
In this case, the ellipses with major and minor axes are parallel to the crystallogrophic $a$ and $b$ axes.

In the present study, we adopt two different approaches. One approach is to directly solve the eigenequation in Eq.~(\ref{eq:H-Mw}) by restricting the number of photon as $|m|\le 16$ where the size of the Floquet Hamiltonian matrix $\mathcal{H}_{nm}(\equiv H_{n-m}-m\omega\delta_{n,m})$ in Eq.~(\ref{eq:H-Mw2}) is 132$\times$132 because $\hat{H}_n$ is a 4$\times$4 matrix. The other approach is to solve the eigenequation for $H_{\rm eff}$ in Eq.~(\ref{eq:Heff}) by restricting the summation over $n$ within $n \le 16$. We have confirmed that the results do not alter between these two approaches even quantitatively except for a few points in the phase diagrams. The former method provides finer and more precise structures of the phase diagrams particularly in the low-frequency region of $\hbar\omega/t<1$ because the latter method is based on the effective Hamiltonian valid in the high-frequency limit. We also note that for smaller frequencies $\omega$, a larger size $|m|$ of the Floquet matrix is required, typically of the order of $W/\hbar\omega$, where $W$ is the band width~\cite{Mikami16}. Since we adopt $|m|\le 16$ for $\alpha$-(BEDT-TTF)$_2$I$_3$ with $W\sim0.75$ eV, the obtained results are trustable for $\hbar\omega \ge$ 0.05 eV. Thus, in the following, we mainly discuss the results obtained by the former method unless otherwise noted.

\section{4. Calculations of Physical Quantities}
\subsection{4.1 Chern number}
The Chern number of the $\nu$th band $N_{\rm Ch}^\nu$ ($\nu$=1,2,3,4) is related to the Berry curvature $B_z^{\nu}(\bm k)$,
\begin{eqnarray}
N_{\rm Ch}^\nu=\frac{1}{2\pi}\int\int_{\rm BZ}\;B_z^{\nu}(\bm k) dk_xdk_y,
\end{eqnarray}
where the Berry curvature $B_z^{\nu}(\bm k)$ at each $\bm k$ point is given by
\begin{align}
&B_z^{\nu}(\bm k)=
\nonumber \\
&i\sum_{(m,\mu)}\frac{
\bra{\Phi_{\nu}^n(\bm k)}\frac{\partial \hat{\mathcal{H}}}{\partial k_x}\ket{\Phi_{\mu}^m(\bm k)}
\bra{\Phi_{\mu}^m(\bm k)}\frac{\partial \hat{\mathcal{H}}}{\partial k_y}\ket{\Phi_{\nu}^n(\bm k)}
-{c.c.}}
{[\varepsilon^m_\mu(\bm k)-\varepsilon^n_\nu(\bm k)]^2}.
\end{align}
Here $\hat{\mathcal{H}}$ denotes the Floquet Hamiltonian matrix, whereas $\varepsilon^n_\nu(\bm k)$ and $\ket{\Phi_{\nu}^n(\bm k)}$ are, respectively, the eigenenergies and the corresponding eigenvectors of Eq.~(\ref{eq:H-Mw}) with $\nu=1,2,3,4$ and $|m|\le 16$. The summation is taken over $m$ and $\mu$ where $(m,\mu)\ne(n,\nu)$; ``$c.c.$" denotes the complex conjugate of the first term of the numerator. Note that $B_z^{\nu}$ is independent of the photon number $n$. In this work, the Chern numbers are calculated using a numerical technique proposed by Fukui $et$ $al.$ in Ref.~\cite{Fukui05}.

\subsection{4.2 Hall conductivity}
The Hall conductivity $\sigma_{xy}$ is a physical quantity sensitive to the topological properties of electronic states and thus can be exploited to identify topological phases. This quantity is associated with the Berry curvatures of bands $B_z^{\nu}(\bm k)$ as
\begin{eqnarray}
\sigma_{xy}=\frac{2e^2}{h}\int\int_{\rm BZ}\;\frac{dk_xdk_y}{2\pi}
\sum_{\nu} n_{\nu}(\bm k) B_z^{\nu}(\bm k).
\label{sgmyx}
\end{eqnarray}
The factor 2 comes from the spin degeneracy. Here $n_{\nu}(\bm k)$ is the nonequilibrium distribution function for the $\nu$-th Floquet band in the photodriven steady states~\cite{Aoki14,Tsuji08,Tsuji09},
\begin{eqnarray}
n_{\nu}(\bm k)=
\frac{\braket{\Phi_\nu^0(\bm k)|\hat{N}_{\bm k}(\varepsilon_\nu^0(\bm k))|\Phi_\nu^0(\bm k)}}
{\braket{\Phi_\nu^0(\bm k)|\hat{A}_{\bm k}(\varepsilon_\nu^0(\bm k))|\Phi_\nu^0(\bm k)}}.
\label{eq:occupation}
\end{eqnarray}
This quantity is calculated using the Floquet--Keldysh formalism~\cite{Aoki14,Tsuji08,Tsuji09}, which combines the Keldysh Green's function technique~\cite{Jauho94,Mahan00} with the Floquet theory. The quantities $\hat{A}_{\bm k}$ and $\hat{N}_{\bm k}$ are given by,
\begin{eqnarray}
& &\hat{A}_{\bm k}(\varepsilon)=\frac{i}{2\pi} \left(
\hat{G}^{\rm R}(\bm k,\varepsilon)-\hat{G}^{\rm A}(\bm k,\varepsilon)\right),
\\
& &\hat{N}_{\bm k}(\varepsilon)=-\frac{i}{2\pi}\hat{G}^{<}(\bm k,\varepsilon).
\end{eqnarray}
The lesser Green's function $\hat{G}^{<}$ and the lesser self-energy $\hat{\Sigma}^{<}$ are calculated, respectively, by,
\begin{eqnarray}
& &\hat{G}^{<}(\bm k,\varepsilon)=\hat{G}^{\rm R}(\bm k,\varepsilon)
\;\hat{\Sigma}^{<}(\varepsilon)\;\hat{G}^{\rm A}(\bm k,\varepsilon), \\
& &\hat{\Sigma}^{<}(\varepsilon)=\frac{
\hat{\Sigma}^{\rm A}+\hat{\Sigma}^{\rm K}(\varepsilon)-\hat{\Sigma}^{\rm R}}{2}.
\end{eqnarray}
The retarded, advanced and Keldysh Green's functions $\hat{G}^{\rm R}$, $\hat{G}^{\rm A}$, and $\hat{G}^{\rm K}$ are calculated by solving the following Dyson equation,
\begin{eqnarray}
& &\left(
\begin{array}{cc}
\hat{G}^{\rm R}(\bm k,\varepsilon) & \hat{G}^{\rm K}(\bm k,\varepsilon) \\
0                                    & \hat{G}^{\rm A}(\bm k,\varepsilon)
\end{array}
\right)^{-1}
\nonumber \\
& &=
\left(
\begin{array}{cc}
[\hat{G}^{\rm R0}(\bm k,\varepsilon)]^{-1} & 0 \\
0 & [\hat{G}^{\rm A0}(\bm k,\varepsilon)]^{-1}
\end{array}
\right)
-\left(
\begin{array}{cc}
\hat{\Sigma}^{\rm R} & \hat{\Sigma}^{\rm K}(\varepsilon) \\
0 & \hat{\Sigma}^{\rm A}
\end{array}
\right),
\nonumber \\
\end{eqnarray}
where $\hat{\Sigma}^{\rm R}$, $\hat{\Sigma}^{\rm A}$ and $\hat{\Sigma}^{\rm K}$ are matrices of the retarded, advanced, and Keldysh self-energies, respectively. Each of the matrices is composed of $(2m_{\rm max}+1)\times(2m_{\rm max}+1)$ block matrices where we set $m_{\rm max}=16$ in the present study, while the size of each block matrix is 4$\times$4. The matrix components of $\hat{G}^{\rm R0}$, $\hat{G}^{\rm A0}$, $\hat{\Sigma}^{\rm R}$, $\hat{\Sigma}^{\rm A}$ and $\hat{\Sigma}^{\rm K}$ are given respectively by
\begin{align}
&[\hat{G}^{\rm R0}(\bm k,\varepsilon)]^{-1}_{n\nu,m\mu}
=\varepsilon\delta_{n,m}\delta_{\nu,\mu}-\mathcal{H}_{n\nu,m\mu}(\bm k),
\\
&[\hat{G}^{\rm A0}(\bm k,\varepsilon)]^{-1}_{n\nu,m\mu}
=\varepsilon\delta_{n,m}\delta_{\nu,\mu}-\mathcal{H}_{n\nu,m\mu}(\bm k),
\\
&[\hat{\Sigma}^{\rm R}]_{n\nu,m\mu}=-i\Gamma\delta_{n,m}\delta_{\nu,\mu},
\\
&[\hat{\Sigma}^{\rm A}]_{n\nu,m\mu}= i\Gamma\delta_{n,m}\delta_{\nu,\mu},
\\
&[\hat{\Sigma}^{\rm K}(\varepsilon)]_{n\nu,m\mu}=-2i\Gamma\tanh\left[
\frac{\varepsilon-\mu+m\omega}{2k_{\rm B}T_{\rm hr}} \right]\delta_{n,m}\delta_{\nu,\mu}.
\nonumber \\
\end{align}
Here the symbol $\hat{M}_{n\nu,m\mu}$ denotes the $(\nu,\mu)$th component of the $(m,n)$th block matrix $\hat{M}_{nm}$ ($4 \times 4$), and the block matrix $\hat{\mathcal{H}}_{n,m}$ constituting the Floquet Hamiltonian is given by Eq.~(\ref{eq:H-Mw2}). We assume that the system is coupled to a heat reservoir at temperature $T_{\rm hr}$ with a dissipation coefficient $\Gamma$ where we set $\Gamma$=0.1 eV for the calculations. For simplicity, we neglect the frequency-dependence and the momentum-dependence of the dissipation coefficient $\Gamma$, for which the dissipations appear only in the diagonal components of the self-energy matrices.

\subsection{4.3 Chiral edge states}
\begin{figure}[b]
\includegraphics[scale=1.0]{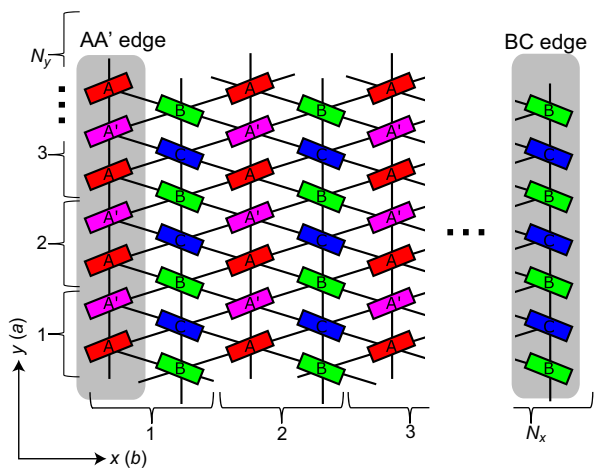}
\caption{Monolayer nanoribbon-shaped system of $\alpha$-(BEDT-TTF)$_2$I$_3$ having AA$^{\prime}$ and BC edges parallel to the $y$ ($a$) axis used for calculation of the band dispersion relations including those of the edge states. The open boundary conditions are imposed in the $x$ direction, whereas the periodic boundary conditions in the $y$ direction. The numberings of the unit cells along the $x$ and $y$ axes are also presented.}
\label{Fig02}
\end{figure}
In addition to the Chern numbers, the emergence of conductive chiral edge states is another feature of the topologically nontrivial Chern insulator phase. The band dispersion relations including those associated with the edge states are calculated for systems having edges by imposing the open boundary conditions in the $x$ direction and the periodic boundary conditions in the $y$ direction and vice versa. Here we explain how to perform the calculations by taking a system having AA$^{\prime}$ and BC edges, both of which are parallel to the $y$ axis (see Fig.~\ref{Fig02}). The positions of unit cells are numbered in ascending order from the left most ($i$=1) to the right most ($i=N_x$) along the $x$ axis, whereas from the bottom ($j$=1) to the top ($j=N_y$) along the $y$ axis.
Because we impose the periodic boundary conditions in the $y$ direction, we can perform the Fourier transformation for the creation and annihilation operators with respect to the $y$ coordinate, 
\begin{eqnarray}
c^{\dagger}_{ij\alpha}=\frac{1}{\sqrt{N_y}}\sum_{k_y}
c^{\dagger}_{k_yi\alpha}e^{ik_y y_{j,\alpha}},
\end{eqnarray}
where $i$ and $j$ are the integer coordinates of unit cells along the $x$ and $y$ axes (i.e., $1\le i \le N_x$ and $1\le j \le N_y$), respectively. The index $\alpha$ specifies a molecule site among four in the unit cell ($\alpha$=1,2,3,4). The coordinate $y_{i,\alpha}$ is the $y$ coordinate of the $\alpha$th molecule site in the (1,$j$)th unit cell. Then we can solve the eigenvalue problem for the nanoribbon-shaped system shown in Fig.~\ref{Fig02} using the basis $\left(c_{k_y,A,1}^{\dagger}, c_{k_y,A^{\prime},1}^{\dagger}, \cdots , c_{k_y,B,N_x}^{\dagger}, c_{k_y,C,N_x}^{\dagger}\right)$ with 
\begin{eqnarray}
	k_y = \frac{2\pi}{N_ya}n_y, \left(n_y = -\frac{N_y}{2}+1, \cdots, \frac{N_y}{2}\right).
\end{eqnarray}

\section{5. Results}
\subsection{5.1 Phase diagrams}
\begin{figure*}[tbh]
\includegraphics[scale=0.5]{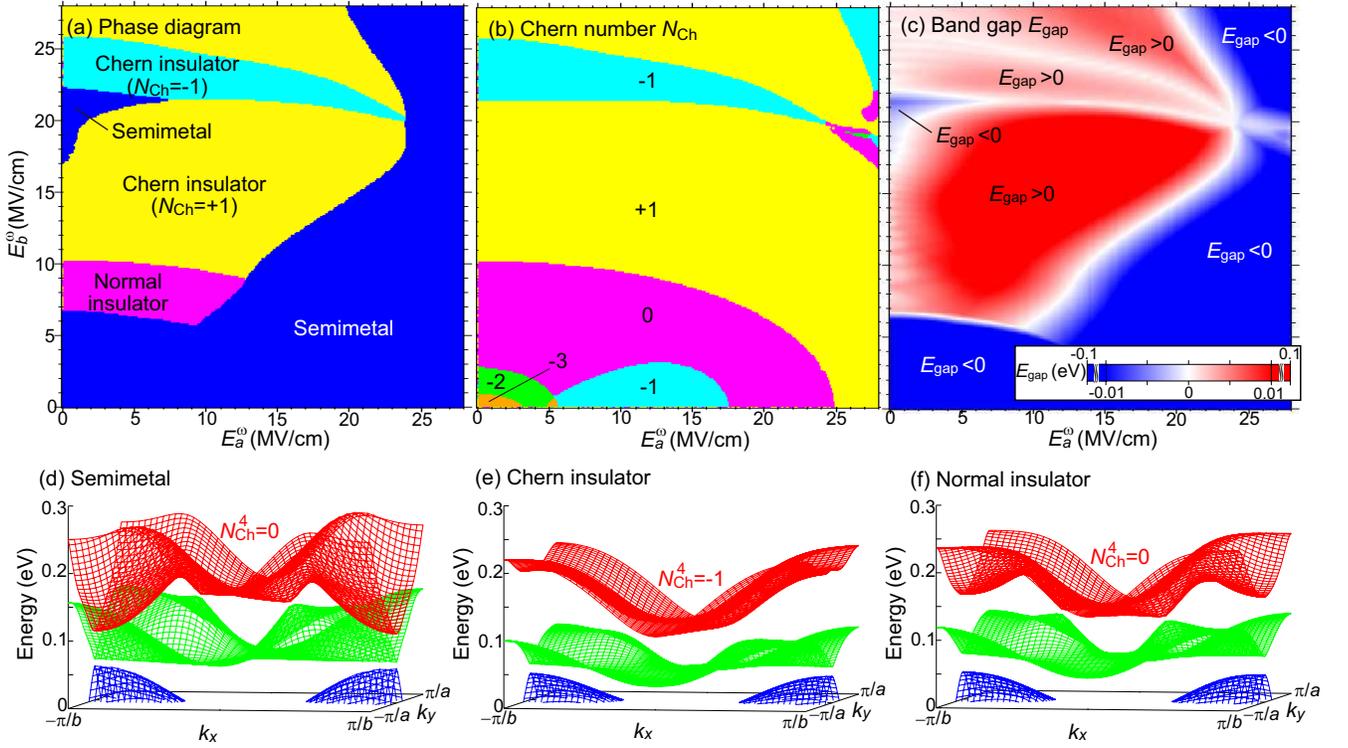}
\caption{(a) Phase diagram of photodriven $\alpha$-(BEDT-TTF)$_2$I$_3$ in the plane of the electric-field amplitudes $E_a^\omega$ and $E_b^\omega$ of applied elliptically polarized light. (b, c) Color maps of (b) the Chern number of the highest (fourth) band $N_{\rm Ch}^4$ and (c) the band gap $E_{\rm gap}$ in the plane of $E_a^\omega$ and $E_b^\omega$. (d-f) Typical band structures for (d) the Floquet semimetal phase ($E_a^\omega=8$ MV/cm and $E_b^\omega=3$ MV/cm), (e) the Floquet Chern insulator phase ($E_a^\omega=11$ MV/cm and $E_b^\omega=11$ MV/cm), and (f) the Floquet normal insulator phase ($E_a^\omega=5$ MV/cm and $E_b^\omega=8$ MV/cm). The light frequency is fixed at $\hbar\omega=0.5$ eV.}
\label{Fig03}
\end{figure*}
\begin{figure*}[tbh]
\includegraphics[scale=0.5]{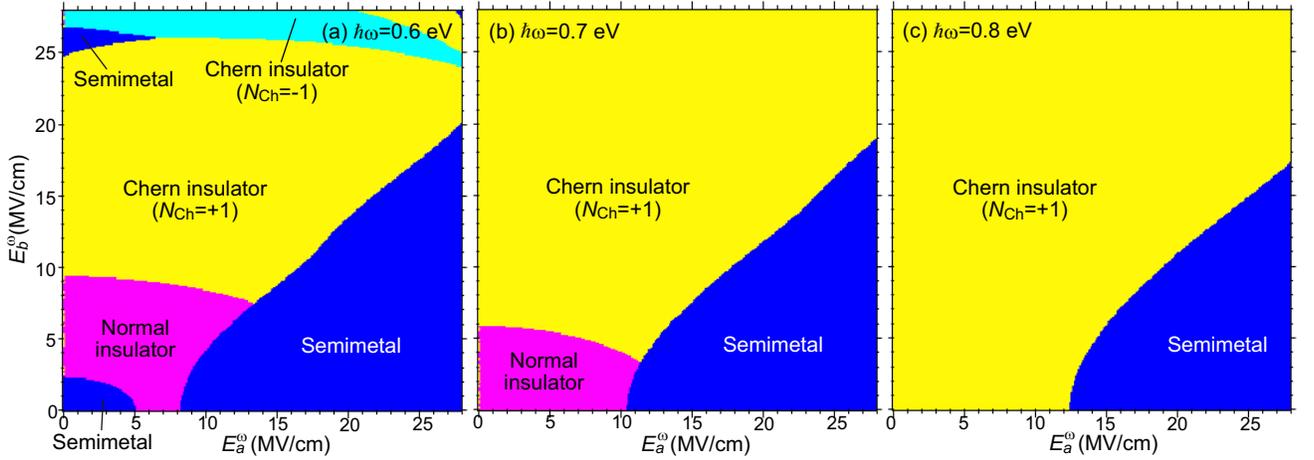}
\caption{Phase diagrams of photodriven $\alpha$-(BEDT-TTF)$_2$I$_3$ in the plane of $E_a^\omega$ and $E_b^\omega$ of applied elliptically polarized light for various light frequencies, (a) $\hbar\omega=0.6$ eV, (b) $\hbar\omega=0.7$ eV, and (c) $\hbar\omega=0.8$ eV.}
\label{Fig04}
\end{figure*}
We first discuss a nonequilibrium phase diagram of photodriven $\alpha$-(BEDT-TTF)$_2$I$_3$ under irradiation with elliptically polarized light. Because the crystal structure of $\alpha$-(BEDT-TTF)$_2$I$_3$ is not symmetric with respect to the interchange of the $a$ and $b$ axes, a rich phase diagram is obtained in the plane of electric-field amplitudes $E_a$ and $E_b$ for the elliptically polarized light. Figure~\ref{Fig03}(a) presents a phase diagram in the plane of $E_a^\omega$ and $E_b^\omega$ when the light frequency is fixed at $\hbar\omega=0.5$ eV. The phase diagram contains three phases, that is, the Floquet semimetal phase, the Floquet Chern insulator phase, and the Floquet normal insulator phase. These phases are classified by calculating the Chern number $N_{\rm Ch}$ [Fig.~\ref{Fig03}(b)] and the band gap $E_{\rm gap}$ [Fig.~\ref{Fig03}(c)] as well as the quasienergy band structures [Figs.~\ref{Fig03}(d)-(f)]. The band gap $E_{\rm gap}$ is defined by,
\begin{equation}
E_{\rm gap} = \min[\varepsilon_4^0(\bm{k})] - \max[\varepsilon_3^0(\bm{k})],
\end{equation}
where $\min[\varepsilon_4^0(\bm{k})]$ is the minimum quasienergy of the fourth band, while $\max[\varepsilon_3^0(\bm{k})]$ is the maximum quasienergy of the third band in the zero-photon subspace (i.e., $n$=0). When $E_{\rm gap}$ is negative (positive), the phase is classified into the Floquet semimetal (insulator) phase. We further classify the insulator phases into two kinds, that is, the topologically non-trivial Floquet Chern insulator phase and the topologically trivial Floquet normal insulator phase according to the Chern number $N_{\rm Ch}$, which is defined as a sum of the Chern numbers of bands below the Fermi level,
\begin{equation}
N_{\rm Ch}= \sum_{\nu=1}^3 N_{\rm Ch}^\nu = -N_{\rm Ch}^4.
\end{equation}
Namely, the band structures of the photoinduced Floquet phases in Figs.~\ref{Fig03}(d)-(f) are respectively characterized by $E_{\rm gap}$ and $N_{\rm Ch}$ as summarized in Table~\ref{tab:phsd}.
\begin{table}[tb]
\begin{tabular}{|c||c|c|}
\hline
Floquet phase & Band gap $E_{\rm gap}$ & Chern number $N_{\rm Ch}$ \\
\hline \hline
Semimetal & $E_{\rm gap}\le0$ & $N_{\rm Ch}=0$ \\ 
\hline
Chern insulator & $E_{\rm gap}>0$ & $N_{\rm Ch}\ne0$ \\ 
\hline
Normal insulator & $E_{\rm gap}>0$ & $N_{\rm Ch}=0$ \\ 
\hline
\end{tabular}
\caption{Classification of the photoinduced phases in $\alpha$-(BEDT-TTF)$_2$I$_3$ according to the band gap $E_{\rm gap}$ and the Chern number $N_{\rm Ch}$.}
\label{tab:phsd}
\end{table}

As explained in Sect.~3, the present calculations are performed using the Floquet Hamiltonian matrix $\mathcal{H}_{nm}$ in Eq.~(\ref{eq:H-Mw2}) after we truncate the matrix size to $|m|\le16$, instead of the effective Floquet Hamiltonian $H_{\rm eff}$ in Eq.~(\ref{eq:Heff}). Note that we obtain the Floquet normal insulator phase characterized by $E_{\rm gap}>0$ and $N_{\rm Ch}=0$ when both amplitudes (i.e., $E_a^\omega$ and $E_b^\omega$) and frequency $\hbar\omega$ are low. This phase, however, cannot be reproduced by calculations with the effective Floquet Hamiltonian $H_{\rm eff}$, which indicates that the Floquet normal insulator phase is realized by effects which are not incorporated in the $1/\omega$ expansion in the high-frequency limit.

Figures~\ref{Fig04}(a)-(c) present obtained nonequilibrium phase diagrams of photodriven $\alpha$-(BEDT-TTF)$_2$I$_3$ for various light frequencies of (a) $\hbar\omega=0.6$ eV, (b) $\hbar\omega=0.7$ eV, and (c) $\hbar\omega=0.8$ eV. In addition to these frequency points, we have studied the phase diagrams at several frequency points although they are not presented here. We find that the Floquet normal insulator phase appears when $\hbar\omega\lesssim 0.75$ eV, but its area in the phase diagram gets smaller as the light frequency $\hbar\omega$ increases and eventually disappears for $\hbar\omega\gtrsim 0.75$ eV. These results indicate that we can selectively realize Floquet phases even with rather weak light amplitudes by tuning the light frequency $\hbar\omega$. Namely, we can have an opportunity to observe the Floquet semimetal phase for $\hbar\omega\sim 0.6$ eV, the Floquet normal insulator phase for $\hbar\omega\sim 0.7$ eV, and the Floquet Chern insulator phase for $\hbar\omega\sim 0.8$ eV with a rather weak light fields $E_a^\omega$ and $E_b^\omega$ even less than 1 MV/cm. This fact increases the experimental feasibility, and we expect that the phase diagrams obtained here are helpful for designing the experiments.

In the present study, we consider a situation with sufficiently small dissipations where the system can reach a nonequilibrium steady state which evolves time-periodically synchronizing with the ac electric fields of light even when the amplitudes $E_a^\omega$ and $E_b^\omega$ are infinitesimally small. In this ideal situation, our theoretical treatment based on the Floquet theorem is justified, and the obtained phase diagrams are valid even in the limit of $E_a^\omega \rightarrow 0$ and $E_b^\omega \rightarrow 0$, where infinitesimally small gaps are opened at the Dirac points and the system becomes the insulator or semimetal. (Note that the point at $E_a^\omega=0$ and $E_b^\omega=0$ is a singular point where the gaps are exactly closed). On the contrary, the dissipation effect necessarily exists in real materials which hinders the realization of nonequilibrium steady state when the light electric fields are weak and thus makes the Floquet description inappropriate. Therefore, the areas around $E_a^\omega\sim 0$ and $E_b^\omega\sim 0$ in the present phase diagrams must be modified when we take the dissipation effect into account. We, however, expect that overall features of the phase diagram hold even in real systems with finite dissipations. The effect of dissipations on the threshold amplitude of light electric fields for the predicted photoinduced phase transitions should be clarified in the future by a more elaborate theory. 

\subsection{5.2 Chiral edge states}
\begin{figure*}[tbh]
\includegraphics[scale=0.5]{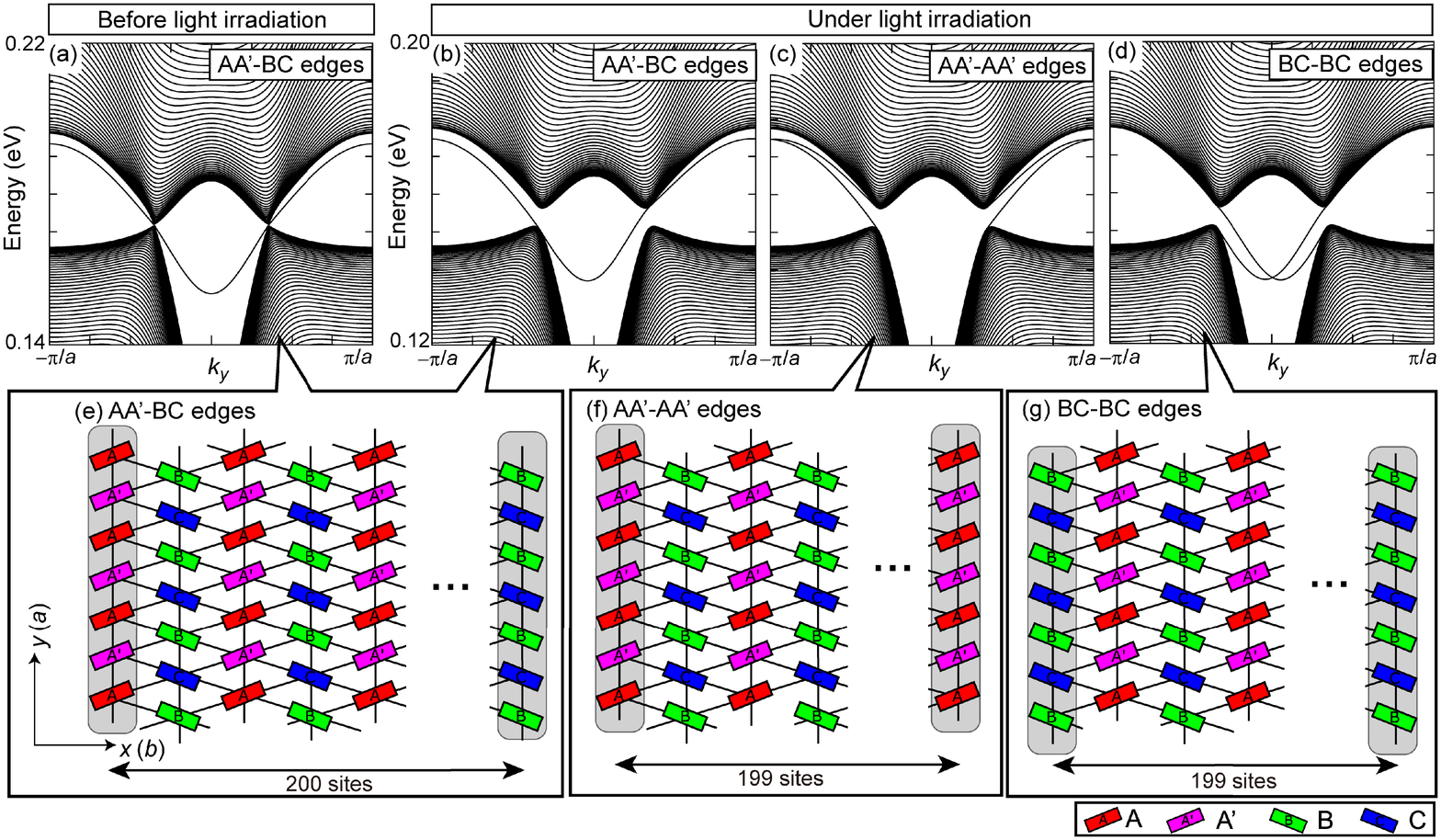}
\caption{Quasienergy band structures with chiral edge-state bands for systems with vertical edges on which the open boundary conditions are imposed in the $x$ ($b$) direction. (a) Those for a steady system with AA$^{\prime}$-BC edges before light irradiation. (b-d) Those for driven systems with (b) AA$^{\prime}$-BC edges (c) AA$^{\prime}$-BC edges, and (d) BC-BC edges where the periodic boundary conditions are imposed in the $y$ ($a$) direction. (e-f) Systems with (e) AA$^{\prime}$-BC edges, (f) AA$^{\prime}$-AA$^{\prime}$ edges, and (g) BC-BC edges used for the calculations. The amplitudes and frequency of light are chosen to be $E_a^\omega=8$ MV/cm, $E_b^\omega=8$ MV/cm, and $\hbar\omega=0.7$ eV.}
\label{Fig05}
\end{figure*}
\begin{figure*}[tbh]
\includegraphics[scale=0.5]{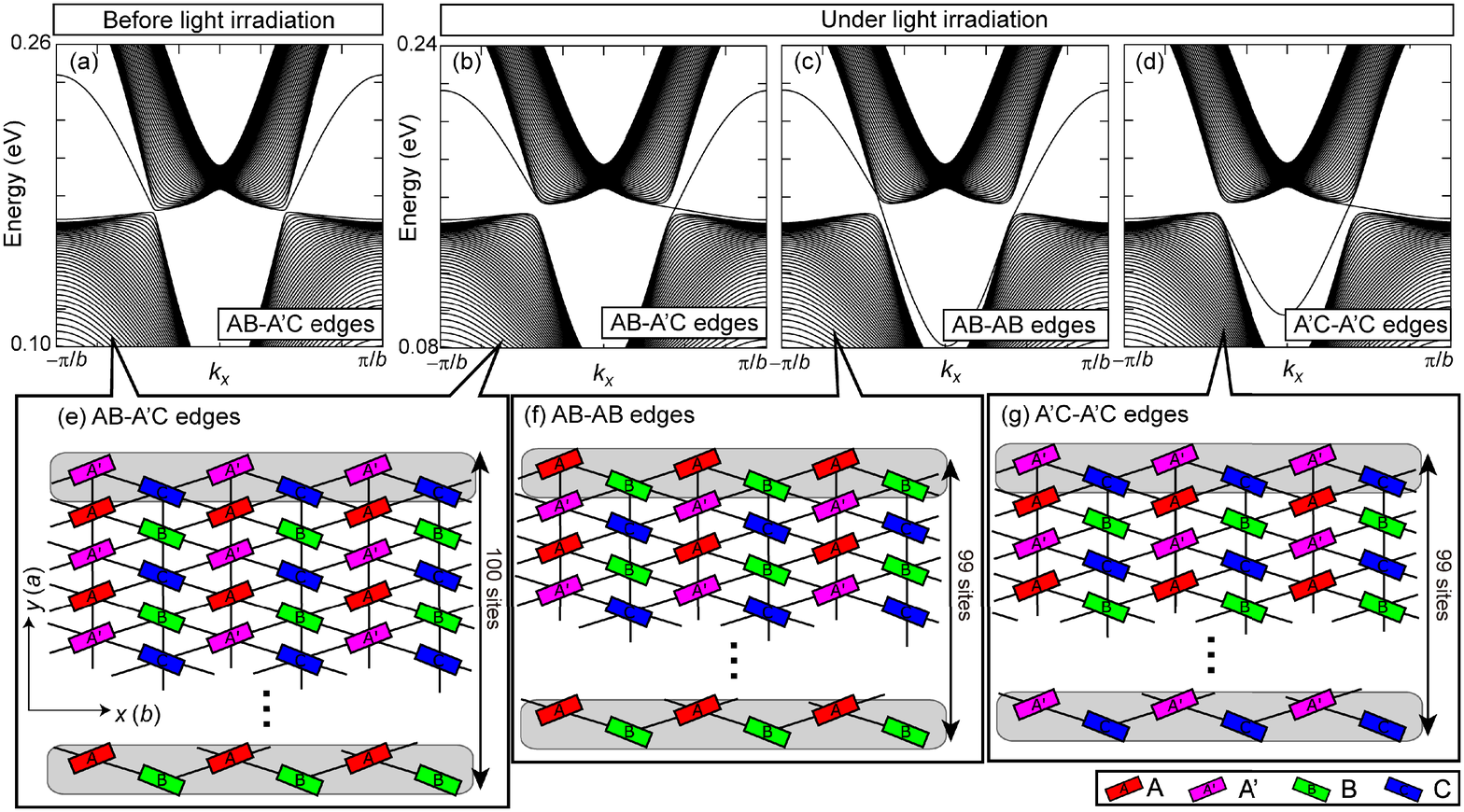}
\caption{Quasienergy band structures with chiral edge-state bands for systems with horizontal edges on which the open boundary conditions are imposed in the $y$ ($a$) direction. (a) Those for a steady system with AB-A$^{\prime}$C edges before light irradiation. (b-d) Those for driven systems with (b) AB-A$^{\prime}$C edges (c) AB-AB edges, and (d) A$^{\prime}$C-A$^{\prime}$C edges where the periodic boundary conditions are imposed in the $x$ ($b$) direction. (e-f) Systems with (e) AB-A$^{\prime}$C edges, (f) AB-AB edges, and (g) A$^{\prime}$C-A$^{\prime}$C edges used for the calculations. The amplitudes and frequency of light are chosen to be $E_a^\omega=8$ MV/cm, $E_b^\omega=8$ MV/cm, and $\hbar\omega=0.7$ eV.}
\label{Fig06}
\end{figure*}
The Chern insulator is characterized by existence of chiral edge states as well as the Chern number $N_{\rm Ch}$. Therefore, we next investigate the edge states for systems having edges to validate the predicted photoinduced Chern insulator phase. We examine several monolayer systems of $\alpha$-(BEDT-TTF)$_2$I$_3$ with different edges. Figures~\ref{Fig05}(a) and (b) present the calculated band dispersion relations before and during the light irradiation, respectively, for a system with vertical edges along the $y$ axis, that is, the AA$^{\prime}$ and BC edges at left and right ends of the sample [Fig.~\ref{Fig02} and Fig.~\ref{Fig05}(e)]. As explained in Sect.~4, the open boundary conditions are imposed in the $x$ direction, while the periodic boundary conditions are imposed in the $y$ direction. Note that we obtain the band dispersion relations including those associated with the edge states by diagonalizing the Floquet effective Hamiltonian. Here the amplitudes and frequency of the elliptically polarized light are chosen to be $E_a^\omega=8$ MV/cm, $E_b^\omega=8$ MV/cm, and $\hbar\omega=0.7$ eV, respectively, where the Chern insulator is realized. The band dispersion relations in Figs.~\ref{Fig05}(a) and (b) show appearance of the bands associated with the edge states. The Dirac points are degenerate before light irradiation in Fig.~\ref{Fig05}(a), whereas they become gapped under light irradiation in Fig.~\ref{Fig05}(b). Importantly, the edge-state bands connect the lower valence band and the upper conduction band, which clearly evidences that the predicted photoinduced phase assigned to the Chern insulator phase is indeed topologically nontrivial. We also examine other systems with different vertical edges. Figure~\ref{Fig05}(c) presents the band dispersion relations for a system having two AA$^{\prime}$ edges [Fig.~\ref{Fig05}(f)], whereas Fig.~\ref{Fig05}(d) presents those for a system having two BC edges [Fig.~\ref{Fig05}(g)]. We again obtain the band dispersions associated with the chiral edge states connecting the upper and lower bands of the gapped Dirac cones.

We also calculate the band dispersion relations for systems with horizontal edges [see Figs.~\ref{Fig06}(a)-(d)]. Systems used for the calculations are presented in Figs.~\ref{Fig06}(e)-(g). Specifically, we examine the systems with (e) AB-A$^\prime$C edges, (f) AB-AB edges, and (g) A$^{\prime}$C-A$^{\prime}$C edges at upper and bottom ends of the system. The open boundary conditions are imposed in the $y$ direction, whereas the periodic boundary conditions are imposed in the $x$ direction. The amplitudes and frequency of light are again set to be $E_a^\omega=8$ MV/cm, $E_b^\omega=8$ MV/cm, and $\hbar\omega=0.7$ eV, respectively. The Dirac points are degenerate before the light irradiation [Fig.~\ref{Fig06}(a)], but the degeneracy is lifted by the photoinduced gap during the light irradiation [Figs.~\ref{Fig06}(b)-(d)]. Although the band dispersions associated with the edge states depend on the edge species, they always connect the lower valence band and the upper conduction band separated by the photoinduced band gap at the Dirac points, again indicating that the corresponding photoinduced Chern insulator phase is indeed topologically nontrivial.

\subsection{5.3 Hall conductivities}
\begin{figure*} 
\includegraphics[scale=0.5]{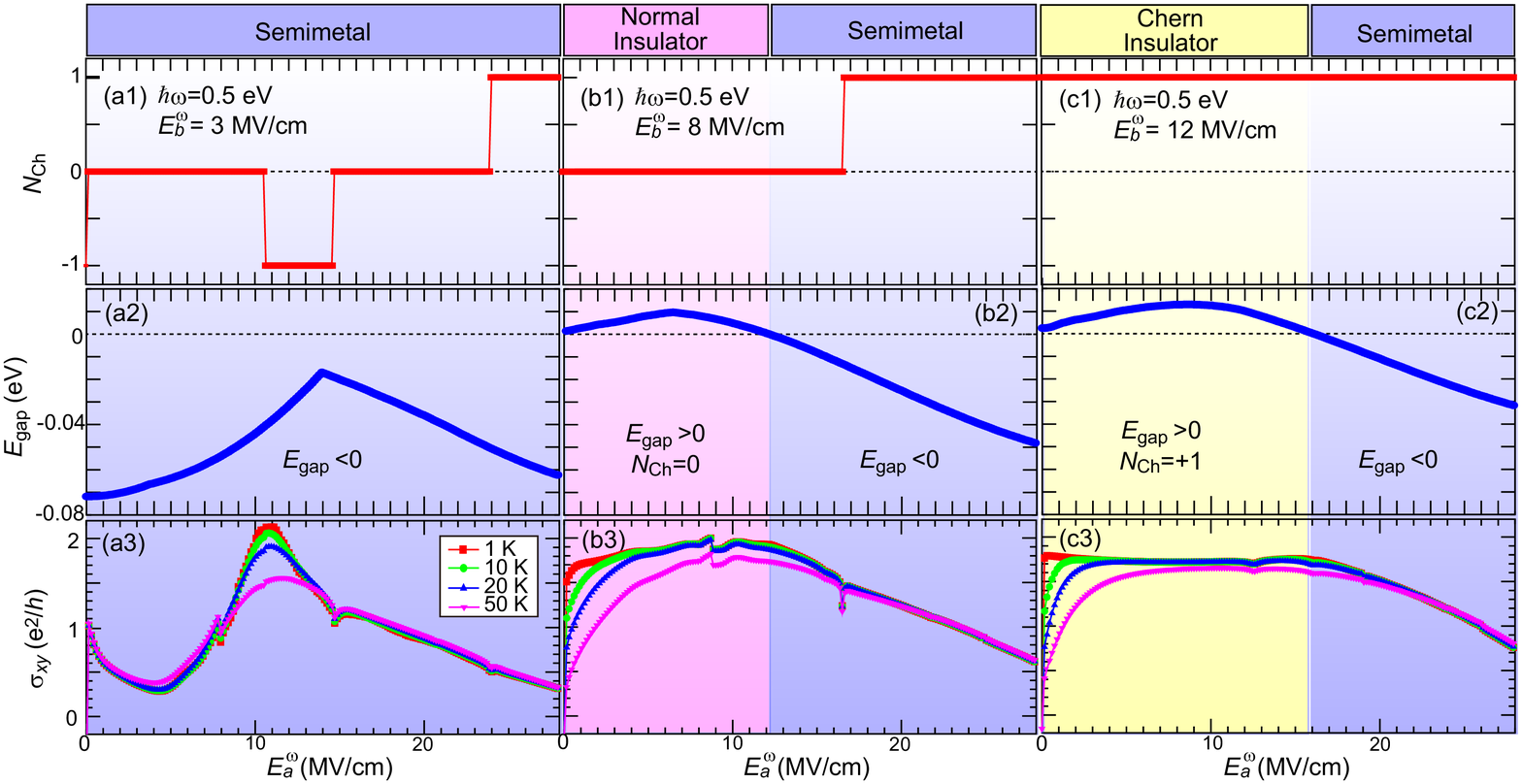}
\caption{Calculated $E_a^\omega$ dependencies of physical quantities for a fixed light frequency of $\hbar\omega$=0.5 eV. (a1)-(a3) Chern number $N_{\rm Ch}$ (a1), band gap $E_{\rm gap}$ (a2), and Hall conductivity $\sigma_{xy}$ (a3) for $\alpha$-(BEDT-TTF)$_2$I$_3$ irradiated with elliptically polarized light as functions of $E_a^\omega$ when $E_b^\omega=3$ MV/cm, for which the system is always in the semimetallic phase. (b1)-(b3) Those when $E_b^\omega=8$ MV/cm, for which the system exhibits a phase transition from the normal insulator phase to the semimetallic phase. (c1)-(c3) Those when $E_b^\omega=12$ MV/cm, for which the system exhibits a phase transition from the Chern insulator phase with $N_{\rm Ch}=+1$ to the semimetallic phase. Note that the calculations are hypothetically performed even for unrealistically large light amplitudes $E_a^\omega$ to discuss inherent behaviors of the physical quantities.}
\label{Fig07}
\end{figure*}
\begin{figure*} 
\includegraphics[scale=0.5]{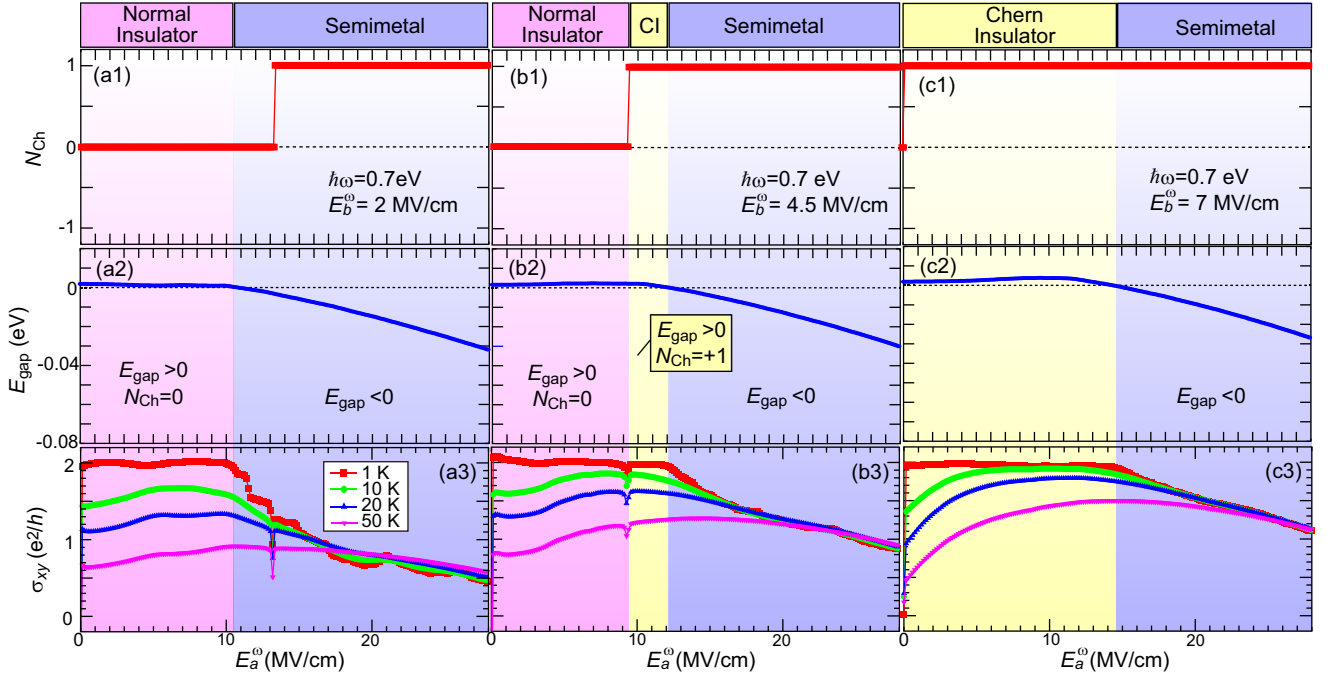}
\caption{Calculated $E_a^\omega$ dependencies of physical quantities for a fixed light frequency of $\hbar\omega$=0.7 eV. (a1)-(a3) Chern number $N_{\rm Ch}$ (a1), band gap $E_{\rm gap}$ (a2), and Hall conductivity $\sigma_{xy}$ (a3) for $\alpha$-(BEDT-TTF)$_2$I$_3$ irradiated with elliptically polarized light as functions of $E_a^\omega$ when $E_b^\omega=2$ MV/cm, for which the system exhibits a phase transition from the normal insulator phase to the semimetallic phase. (b1)-(b3) Those when $E_b^\omega=4.5$ MV/cm, for which the system exhibits two phase transitions from the normal insulator phase to the Chern insulator phase with $N_{\rm Ch}=+1$ to the semimetallic phase. (c1)-(c3) Those when $E_b^\omega=7$ MV/cm, for which the system exhibits a phase transition from the Chern insulator phase with $N_{\rm Ch}=+1$ to the semimetallic phase. Note that the calculations are hypothetically performed even for unrealistically large light amplitudes $E_a^\omega$ to discuss inherent behaviors of the physical quantities.}
\label{Fig08}
\end{figure*}
\begin{figure*} 
\includegraphics[scale=0.5]{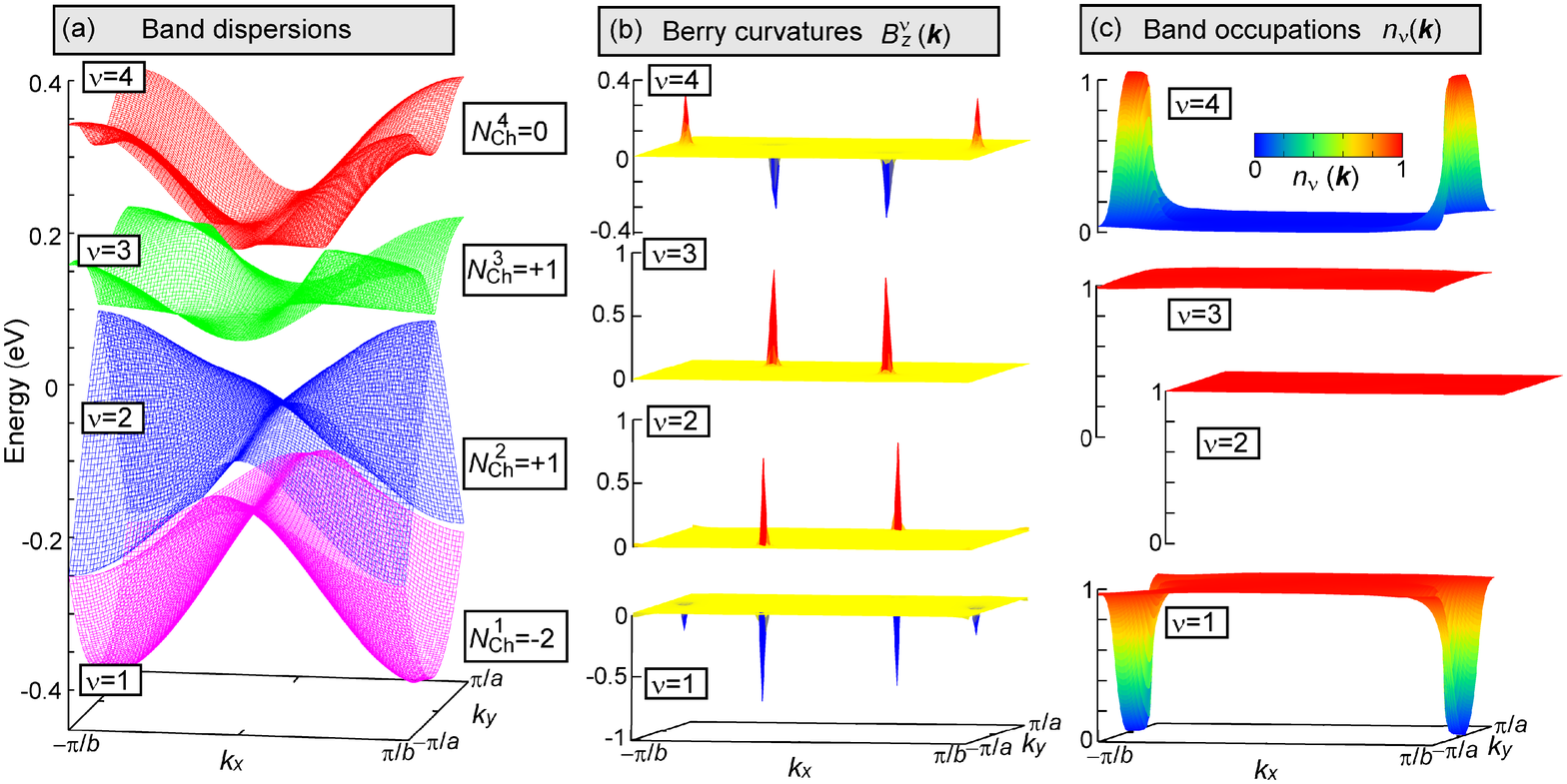}
\caption{Calculated electronic structure in the momentum space for the Floquet normal insulator phase of $\alpha$-(BEDT-TTF)$_2$I$_3$ irradiated with elliptically polarized light when $\hbar\omega=0.7$ eV, $E_a^\omega=5$ MV/cm and $E_b^\omega=4.5$ MV/cm. (a) Floquet band structure in the zero-photon subspace ($n$=0). (b) Berry curvatures $B_z^{\nu}(\bm k)$ of the respective bands with $\nu$=1, 2, 3 and 4. (c) Band occupations $n_{\nu}(\bm k)$ of the respective bands.}
\label{Fig09}
\end{figure*}
We finally study the Hall conductivity in the photoinduced nonequilibrium phases. First, we examine the case where the light frequency is $\hbar\omega=0.5$ eV [see also the phase diagram in Fig.~\ref{Fig03}(a)]. Figure~\ref{Fig07} presents the calculated Chern number $N_{\rm Ch}$, band gap $E_{\rm gap}$, and Hall conductivity $\sigma_{xy}$ as functions of $E_a^\omega$ for several values of $E_b^\omega$, that is, (a1)-(a3) $E_b^\omega=3$ MV/cm, (b1)-(b3) $E_b^\omega=8$ MV/cm, and (c1)-(c3) $E_b^\omega=12$ MV/cm. The Chern number $N_{\rm Ch}$ and the band gap $E_{\rm gap}$ are used to identify the phases. We find that
\begin{itemize}
\item The Floquet Chern insulator phase with $N_{\rm Ch}=1$ has a nearly quantized Hall conductivity $\sigma_{xy}=2e^2/h$ at low temperatures, which is equivalent to the Chern insulator phase in equilibrium.
\item The Floquet normal insulator phase also has a nearly quantized Hall conductivity $\sigma_{xy}=2e^2/h$ despite $N_{\rm Ch}$ is zero.
\item The Floquet semimetal phase has a finite Hall conductivity that gradually decreases as the system becomes distant from the Floquet normal insulator phase [Fig.~\ref{Fig07}(b3)] and the Floquet Chern insulator phase [Fig.~\ref{Fig07}(c3)] with increasing $E_a^\omega$. 
\end{itemize}

Here the second finding is surprising because the quantized Hall conductivity in the Floquet normal insulator phase is in sharp contrast with a vanishing Hall conductivity in the equilibrium normal insulator, which indicates that the Floquet normal insulator phase is not really insulating. This finite and quantized Hall conductivity is observed not only at the small frequency of $\hbar\omega=0.5$ eV, but also at $\hbar\omega=0.7$ eV as seen in Fig.~\ref{Fig08} [see also the phase diagram in Fig.~\ref{Fig04}(b)]. Note that the Floquet normal insulator phase appears only when the light frequency is smaller than 0.75 eV, and hence the frequency of $\hbar\omega=0.7$ eV is almost as high as this threshold frequency. Figures~\ref{Fig08}(a3) and (b3) indeed show that the Hall conductivity in the Floquet normal insulator phase has a quantized value $\sigma_{xy}=2e^2/h$ especially at low temperatures, although they rapidly decrease and the deviation from this quantized value is enhanced as temperature increases because of the small band gap [see Figs.~\ref{Fig08}(a2) and (b2)].

To understand the origin of this quantized Hall conductivity in the Floquet normal insulator phase, we examine the electronic structure of the Floquet normal insulator phase for $\hbar\omega=0.7$ eV, $E_a^\omega=5$ MV/cm and $E_b^\omega=4.5$ MV/cm. Figure~\ref{Fig09}(a) shows the calculated band structure in the zero-photon subspace ($n$=0). The Chern numbers $N^\nu_{\rm Ch}$ for respective bands ($\nu$=1, 2, 3 and 4) are evaluated as $N^1_{\rm Ch}=-2$, $N^2_{\rm Ch}=+1$, $N^3_{\rm Ch}=+1$ and $N^4_{\rm Ch}=0$. The Berry curvatures $B_z^{\nu}(\bm k)$ of the respective bands in the momentum space show momentum-resolved contributions to the Chern number from each band [see Fig.~\ref{Fig09}(b)]. Importantly, the plot of $B_z^{\nu}(\bm k)$ for $\nu$=4 indicates that positive and negative contributions from different momentum points cancel out resulting in the vanishing Chern number of $N^4_{\rm Ch}=0$ in the fourth band. This Berry-curvature structure is a key to understanding the observed quantized Hall conductivity. Figure~\ref{Fig09}(c) shows the nonequilibrium band occupations $n_{\nu}(\bm k)$ for the respective bands calculated using Eq.~(\ref{eq:occupation}). In the equilibrium case where the Fermi-distribution function governs the band occupations, the lower three bands ($\nu$=1,2,3) are fully occupied, while the highest band ($\nu$=4) is almost unoccupied at low temperatures because the electron filling of $\alpha$-(BEDT-TTF)$_2$I$_3$ is 3/4. Thereby, we expect the Chern number $N_{\rm Ch}$ and the Hall conductivity $\sigma_{xy}$ vanish in the equilibrium normal insulator phase. On the other hand, the band occupations in the nonequilibrium system deviate from those in equilibrium governed by the Fermi-distribution function. The topmost panel in Fig.~\ref{Fig09}(c) indicates that the fourth band is partially occupied at the momenta around $(k_x,k_y)=(\pm \pi,0)$ where the Berry curvature $B_z^{4}(\bm k)$ has positive peaks, whereas the bottom panel indicates that the first band is partially unoccupied at the momenta around $(k_x,k_y)=(\pm \pi,0)$ where the Berry curvature $B_z^{1}(\bm k)$ has negative peaks.

The emergence of these positive and negative peaks in the Berry curvature $B_z^{\nu}(\bm k)$ for the $\nu$=4 band as well as the emergence of the Floquet normal insulator phase at lower light frequencies can be understood by the hybridization between the topmost band ($\nu=4$) in the zero-photon subspace ($n=0$) and the bottom band ($\nu=1$) in the one-photon-absorbed subspace ($n=-1$) located above the zero-photon subband set. This band hybridization gives rise to the positive peaks of the Berry curvature at the band crossing points, which offset the negative contributions from the Dirac points on the band labeled with $(n,\nu)=(0,4)$, resulting in the vanishing Chern number and the emergence of the Floquet normal insulator phase. Because the energy spacing between the subband sets of $n=0$ and $n=-1$ is governed by the light frequency $\hbar\omega$, such a band hybridization occurs when the light frequency $\omega$ is small as compared with the bandwidth $W$ (i.e., $\hbar\omega < W$). Conversely, the band hybridization does not occur when $\hbar\omega > W$. Because the bandwidth of $\alpha$-(BEDT-TTF)$_2$I$_3$ in the equilibrium is approximately 0.75 eV, the Floquet normal insulator phase does not appear when $\hbar\omega \geq 0.8$ eV.

\section{6. Summary and Discussion}
We theoretically predicted rich photoinduced phases and phase transitions in organic conductor $\alpha$-(BEDT-TTF)$_2$I$_3$ under irradiation with elliptically polarized light. This compound possesses a pair of tilted Dirac cones in the band structure, and the Dirac points are located on the Fermi level under a uniaxial pressure $P_a$($>2$ kbar). By analyzing a photodriven tight-binding model using the Floquet theory, we obtained rich nonequilibrium phase diagrams in the plane of light amplitudes that contain the Floquet semimetal phase, the Floquet Chern insulator phase, and the Floquet normal insulator phases as nonequilibrium steady states. We demonstrated that all the three phases are accessible with a rather weak light amplitude of less than 1 MV/cm by tuning the light frequency. More specifically, the Floquet semimetal phase, the Floquet normal insulator phase, and the Floquet Chern insulator phase are accessible with weak laser light when $\hbar\omega \sim$0.6 eV, $\hbar\omega \sim$0.7 eV, $\hbar\omega \sim$0.8 eV, respectively. We also revealed that the predicted Floquet Chern insulator phase is characterized by both a quantized topological invariant (i.e., the Chern number) and chiral edge states, which indicates that this phase is indeed topologically nontrivial. In addition, the Floquet normal insulator phase classified according to the band structure and the band Chern numbers turned out to exhibit a quantized Hall conductivity because of partial occupations of the fourth band in the nonequilibrium.

The present organic conductor $\alpha$-(BEDT-TTF)$_2$I$_3$ has several advantages for experimental study on the photoinduced phase-transition phenomena as discussed in Ref.~\cite{Kitayama20}. First, this material is composed of large molecules and thus has relatively large lattice constants ($a$=0.9187 nm and $b$=1.0793 nm)~\cite{Mori12}. These lattice constants are a few times larger than the lattice constant in graphene ($a$=0.246 nm), which effectively enhance the dimensionless laser amplitude $\mathcal{A}_a=eaE^\omega/\hbar\omega$ and $\mathcal{A}_b=ebE^\omega/\hbar\omega$. Second, because the bands associated with the BEDT-TTF layer are well separated from upper and lower band sets by energy gaps~\cite{Kino06}, we have a finite window of light frequency where the BEDT-TTF bands in the zero-photon subspace overlap neither the bands in the one-photon-absorbed ($n=-1$) subspace nor the bands in the one-photon-emitted ($n=+1$) subspace. This is an experimentally suitable situation called off-resonant situation, where the electron occupations of the Floquet bands are well approximated by the Fermi-Dirac distribution function in equilibrium, and, thereby, the photoinduced phases become well-defined.

As we argued above, the present organic compound $\alpha$-(BEDT-TTF)$_2$I$_3$ is an interesting material because we can access various kinds of photoinduced phases with a weak light field by tuning the light frequency. At the same time, our results indicate that richer photoinduced evolutions of the phases can be observed when we apply a much stronger light field. Needless to say, samples of the organic materials might be damaged or even broken under continuous application of an intense light field. Note that it is difficult to evaluate to what extent a sample of this organic compound can endure the intense laser, but still we can argue the feasibility of experiments in the light of recently reported experimental achievements. Some experiments of photoinduced phase transitions have been successfully performed for $\kappa$-type (BEDT-TTF)-based organic salts using a few-cycle-pulse laser light or one-cycle-pulse laser light as intense as $E^\omega=$16 MV/cm~\cite{Uchida15,Uchida16}. In addition, recent experiments demonstrated that a small-number-cycle pulse or even a less-than-10-cycle pulse, instead of continuous-wave photoexcitation, is enough to realize Floquet phases or nonequilibrium steady phases~\cite{Kawakami18,Tachizaki19,Kawakami20}. Considering these experimental achievements, experimental explorations of the predicted photoinduced phases and phase transitions in $\alpha$-(BEDT-TTF)$_2$I$_3$ are feasible or, at least, worth trying. The short duration of light irradiation is also convenient to avoid or suppress the heating effect, although the time scale of heating is uncertain and its theoretical evaluation is important, which is left for a future study.
We expect that the present work widens a range of target materials for research on the photoinduced phase transitions and contribute to development of the optical control of electronic states in matters.

\section{Acknowledgment}
KK is supported by the World-leading Innovative Graduate Study Program for Materials Research, Industry, and Technology (MERIT-WINGS) of the University of Tokyo. MM is supported by JSPS KAKENHI (Grants No. 16H06345, No. 19H00864, No. 19K21858 and No. 20H00337), CREST, the Japan Science and Technology Agency (Grant No. JPMJCR20T1), a Research Grant in the Natural Sciences from the Mitsubishi Foundation, and a Waseda University Grant for Special Research Projects (Project No. 2020C-269 and No. 2021C-566). YT is supported by JSPS KAKENHI (Grants No. 19K23427 and No. 20K03841). MO is supported by JSPS KAKENHI (Grant No. 18H01162).

\end{document}